
\input phyzzx
\overfullrule=0pt

\def\half{{1\over 2}}
\def\for{{1\over 4}}
\def\a{\alpha}

\def\g{\gamma}
\def\d{\delta}
\def\e{\epsilon}

\def\t{\theta}
\def\l{\lambda}
\def\m{\mu}
\def\f{\phi}
\def\n{\nu}

\def\p{\psi}
\def\r{\rho}

\def\x{\chi}

\def\pa {\partial}

\def\xp{x^+}
\def\xm{x^-}
\def\xw{x^1}
\def\xz{x^0}
\def\scripl{{\cal I}_L^+}
\def\scripr{{\cal I}_R^+}
\def\scriml{{\cal I}_L^-}
\def\scrimr{{\cal I}_R^-}

\def\scripml{{\cal I}^{\pm}_L}
\def\scripmr{{\cal I}^{\pm}_R}
\def\tH{'t~Hooft}
\def\smatrix{$S$-matrix}
\def\wosp{{1\over{\sqrt{\pi}}}}
\def\sp{\sqrt{\pi}}

\rightline{SU-ITP-92-16}
\rightline{May 26, 1992}

\vfill

\title{INFORMATION LOSS AND ANOMALOUS SCATTERING\foot{Work supported
in part by NSF grant PHY89-17438}}

\vfill

\author{Amanda Peet\foot{Supported in part by Stanford University
Physics Department Fellowship Fund}\foot{peet@slacvm.bitnet}, Leonard
Susskind and L\'arus Thorlacius\foot{larus@dormouse.stanford.edu}}

\vfill

\address{Physics Department \break Stanford University, Stanford, CA
94305-4060}

\vfill

\abstract
\singlespace
The approach of 't Hooft to the puzzles of black hole evaporation can
be applied to a simpler system with analogous features.  The system
is $1+1$ dimensional electrodynamics in a linear dilaton background.
Analogues of black holes, Hawking radiation and evaporation exist in
this system.  In perturbation theory there appears to be an
information paradox but this gets resolved in the full quantum theory
and there exists an exact $S$-matrix, which is fully unitary and
information conserving.  't Hooft's method gives the leading terms in
a systematic approximation to the exact result.

\vfill\endpage

\REF\hawking{S.~W.~Hawking
\journal Phys. Rev. & D14 (76) 2460.}

\REF\ac{Y.~Aharonov, A.~Casher and S.~Nussinov
\journal Phys. Lett. & 191B (87) 51.}

\REF\steve{S.~B.~Giddings, {\it Black Holes and Massive Remnants},
UCSB preprint, UCSBTH-92-09, hepth@xxx/9203059, March 1992.}

\REF\thooft{G.~'t~Hooft
\journal Nucl. Phys. & B335 (90) 138, and references therein.}

\REF\cghs{C.~G.~Callan, S.~B.~Giddings, J.~A.~Harvey and
A.~Strominger
\journal Phys. Rev. & D 45 (92) R1005.}

\REF\bddo{T.~Banks, A.~Dabholkar, M.~R.~Douglas and M.~O'Loughlin,
{\it Are Horned Particles the Climax of Hawking Evaporation?},
Rutgers University preprint, RU-91-54, hepth@xxx/9201061, January
1992.}

\REF\rst{J.~G.~Russo, L.~Susskind and L.~Thorlacius, {\it Black
Hole Evaporation in 1+1 Dimensions}, Stanford University preprint,
SU-ITP-92-4, hepth@xxx/9201074, January 1992.}

\REF\bghs{B.~Birnir, S.~B.~Giddings, J.~A.~Harvey and A.~Strominger,
{\it Quantum Black Holes}, preprint UCSB-TH-92-08, EFI-92-16,
hepth@xxx/9203042, March 1992.}

\REF\hawkingii{S.~W.~Hawking, {\it Evaporation of Two-Dimensional
Black Holes}, Caltech preprint, CALT-68-1774, hepth@xxx/9203052,
March 1992.}

\REF\lslt{L.~Susskind and L.~Thorlacius, {\it Hawking Radiation and
Back-Reaction}, SU-ITP-92-12, hepth@xxx/9203054, March 1992.}

\REF\strom{A.~Strominger, {\it Fadeev-Popov Ghosts and 1+1
Dimensional Black Hole Evaporation}, UCSB preprint, UCSB-TH-92-18,
hepth@xxx/9205028, May 1992.}

\REF\dealw{S.~P.~de~Alwis, {\it Quantization of a Theory of 2d
Dilaton Gravity}, University of Colorado preprint, COLO-HEP-280,
hepth@xxx/9205069, May 1992.}

\REF\bilcal{A.~Bilal and C.~G.~Callan, {\it Liouville Models of Black
Hole Evaporation}, Princeton University preprint, PUPT-1320,
hepth@xxx/9205089, May 1992.}

\REF\alstrom{M.~Alford and A.~Strominger, {\it S-Wave Scattering of
Charged Fermions by a Black Hole}, preprint NSF-ITP-92-13,
hepth@xxx/9202075, February 1992.}

\REF\cgc{C.~G.~Callan
\journal Phys. Rev. & D25 (82) 2141
\journal Phys. Rev. & D26 (82) 2058
\journal Nucl. Phys. & B212 (83) 391;
C.~G.~Callan and S.~R.~Das
\journal Phys. Rev. Lett. & 51 (83) 1155.}

\REF\rubakov{V.~Rubakov
\journal JETP Lett. & 33 (81) 644.}

\REF\priv{E.~Verlinde and H.~Verlinde, private communication.}

\REF\gidstro{S.~B.~Giddings and A.~Strominger, {\it Dynamics of
Extremal Black Holes}, preprint UCSB-TH-92-01, hepth@xxx/9202004,
February 1992.}

\REF\chrful{S.~M.~Christensen and S.~A.~Fulling
\journal Phys. Rev. & D15 (77) 2088.}

\chapter{Introduction}

It is a controversial issue whether or not quantum coherence can be
maintained
during the formation and subsequent evaporation of a black hole.  At
one end of
the spectrum of opinion is Hawking's suggestion that this process
indicates a
new level of unpredictability introduced into quantum mechanics by
gravity [\hawking].  Another proposal, which is also radical from the
point of view of quantum mechanics, is that information about the
initial quantum state of the
system is carried by a Planck scale stable remnant [\ac,\steve].
Perhaps the most conservative position has been advocated by \tH\ who
argues that this process should be thought of as a conventional
scattering event in which the black hole is an intermediate state
somewhat analogous to a complex intermediate nucleus formed in a
nuclear collision [\thooft].  \tH\ has taken some tentative steps
toward an \smatrix\ description of such events but the precise
meaning of the resulting \smatrix\ remains unclear.
We feel that this approach deserves attention and should be explored.
It may indeed provide a resolution of the above paradox, or else one
would like to see this logical possibility ruled out.

In order to avoid some of the formidable technical obstacles posed by
quantum gravity in $3+1$ dimensions one can instead consider black
hole evolution in $1+1$ dimensions.  Of course this simplified
setting does not capture all the physics of real black holes but it
does contain an information paradox analogous to the one originally
posed by Hawking.  We begin in section~2 by outlining the arguments
leading to \tH 's \smatrix\ in $1+1$ dimensions.  For this discussion
we use a simple model recently proposed by Callan {\it et al.}
[\cghs] and subsequently discussed in [\bddo -\bilcal].  It turns out
that one obtains some exact expressions where approximations had to
be made in the higher dimensional theory.  The physical
interpretation of our \smatrix\ is nevertheless every bit as obscure
as \tH 's.  The main purpose of this paper is to clarify some of the
issues involved by considering a simpler system, which shares many
features with two-dimensional black holes, but can be solved
explicitly.  The system in question is the $1+1$ dimensional
Schwinger model with the unusual feature that the electrodynamic
coupling strength depends on position. It varies from vanishing
coupling at one end of space to infinite coupling at the other.  The
two ends correspond to spatial infinity (weak coupling) and the deep
interior of the black hole (strong coupling).  This is also the
appropriate coupling dependence to describe $s$-wave fermion
scattering off a $3+1$ dimensional extreme magnetic dilaton black
hole[\alstrom].  A similar model arose in the analysis of monopole
catalysis in [\cgc,\rubakov].  The methods we use in this paper may
find application in that context also.

In section~3 we describe the analogy between black hole physics and
$1+1$ dimensional electrodynamics.  The question of the existence of
a unitary \smatrix\ is shown to be similar in the two cases.  In
section~4 we set up and
solve the classical equations for the formation of an object called a
``charge-hole'' by an incoming electric charge.  We then discuss the
electromagnetic analogue of Hawking radiation.  In this section no
attempt is
made to include back-reaction on the electromagnetic field of the
charge-hole due to the emitted radiation.  Section~5 is devoted to
describing \tH 's method as applied to our model and an expression is
derived for an \smatrix\ .  Section~6 uses the method of bosonization
to account for back-reaction and gives an exact expression for the
single-particle elastic \smatrix\ between one-fermion states.  Then
we construct the generalization to arbitrary states and show that the
exact \smatrix\ is a generalization of \tH 's, with well-defined
procedures for extracting amplitudes in Fock space.  Finally, in
section~7, the information problem is briefly discussed for the
electrodynamic and gravitational systems.

\chapter{\tH 's \smatrix\ for $1+1$ dimensional gravity}
In this section we will repeat \tH 's argument for the form of the
quantum
\smatrix\ for black hole physics in a simplified $1+1$ dimensional
context. We will make no attempt in this section to clarify or

interpret 't~Hooft's
theory.  The reader is advised to skim this section lightly and
return
to it after reading the subsequent material.

Consider the following action for $1+1$ dimensional dilaton gravity
[\cghs]:
$$
I =  \int d^2 x \bigl[ e^{-2\f} \bigl( R + 4 (\nabla \f)^2 + 4\l^2
\bigr)
- \half \sum_{i=1}^N \, (\nabla f_i)^2 \bigr] \ .
\eqn\first
$$
This theory has received considerable attention as a toy model for
black hole physics [\cghs -\bilcal] and we will be brief here.  We
use the conformal gauge
$$
g_{++} = g_{--} = 0, \quad g_{+-} = - \half e^{2 \r} \ .
\eqn\second
$$
The linear dilaton vacuum is given in light-cone ``Kruskal"
coordinates by
$$
e^{-2 \r}  = e^{-2\f} = -\l^2 \xp \xm \ ,
\eqn\third
$$
and the classical static black hole solution is
$$
e^{-2 \r}  = e^{-2\f} = - \l^2 \xp \xm + {M \over \lambda} \ .
\eqn\fourth
$$

Let us consider a geometry describing infalling massless matter, in
the form of a shock wave, with energy-momentum tensor
$$
T_{++}^f = {M \over \lambda \xp_0}  \d ( \xp - \xp_0 )
\eqn\fifth
$$
where $\xp_0$ is the coordinate of the null trajectory of the shock
and $M$ is the total energy carried by it.  The
gravitational and dilaton fields are constructed by patching together
the vacuum solution for $\xp < \xp_0$ and a black hole solution with
mass $M$ for $\xp > \xp_0$.  In order to keep the dilaton and metric
continuous at $\xp = \xp_0$, it is necessary to translate the black
hole solution along the $\xm$ axis by $- {M \over \lambda^3 \xp_0}$.
The full solution is
$$
e^{-2 \r}  = e^{-2\f} = - \l^2 \xp \xm - {M \over \lambda \xp_0}(\xp
- \xp_0) \, \t(\xp - \xp_0)  \ .
\eqn\sixth
$$
If further energy $\d M$, in the form of another incoming shock-wave,
is added to the black hole, the result is simply another shift $-{M
\over \lambda^3 \xp_1}$ in $\xm$ on the null trajectory $\xp =
\xp_1$.  This is illustrated in figure~1.  In four dimensions the
corresponding coordinate transformation across the shock front is
more complicated [\thooft].  However, near the horizon and for $\d M
\ll M$ it can be approximated by a simple shift.\foot{That a uniform
shift is the full answer in two dimensions has also been noted by
E.~Verlinde and H.~Verlinde [\priv].}

For a continuous incoming flux $T_{++}(\xp)$ the solution is
$$\eqalign{
e^{-2 \r}  = e^{-2\f} =& - \l^2 \xp \xm - \int_0^{\xp} d \xp_0 \,
T_{++}(\xp_0) \, (\xp - \xp_0)  \cr
=& - \l^2 \xp \xm - P_{+}(\xp) \bigl[ \xp - {1\over P_{+}(\xp)}
\int_0^{\xp} d\xp_0 \, \xp_0 \, T_{++}(\xp_0) \bigr]\> ,  \cr}
\eqn\seventh
$$
where $P_{+}(\xp) = \int_0^{\xp} d\xp_0 \, T_{++}(\xp_0)$ is the
total incoming
Kruskal momentum conjugate to $\xp$.  From this expression it is
clear that the final black hole geometry is indistinguishable at the
classical level from a black hole formed by a single incoming shock
wave carrying energy $\bar M =\lambda \int_0^\infty d\xp_0 \, \xp_0
\,T_{++}(\xp_0)$ in along
$\bar x_0^+ = {\bar M\over \lambda P_{+}(\infty)}$.

In [\thooft] \tH\ argues that such coordinate shifts influence the
quantum vacuum of the matter fields.  In particular, infalling matter
will induce a unitary transformation on the outgoing modes,
$$
U = \exp(i \d \xm P_{-}) \ ,
\eqn\eighth
$$
where $P_{-} = \int_{-\infty}^{0} d \xm_0 \, T_{--}(\xm_0)$ generates
$\xm$ translations of the Kruskal coordinates and $\d \xm =
P_{+}(\infty)$ is the coordinate shift calculated above.  Written in
a more symmetric form \tH 's \smatrix\ is
$$
S = \exp({i\over \lambda^2} P_+ P_-) \ .
\eqn\ninth
$$
The proper interpretation of this expression is elusive.  It should
be pointed
out that \ninth\ cannot be the final answer.  Indeed, the final state
obtained in this way does not reflect any properties of the initial
state except the total incoming Kruskal momentum, so this \smatrix\
cannot keep track of the full structure of quantum states.

In section~5 a similar line of reasoning will lead to an analogous
expression for an \smatrix\ (with the same shortcomings) in our $1+1$
dimensional electrodynamics.  In section~6 we go on to derive a fully
unitary \smatrix\ and show how the \tH -like result is the leading
term in a systematic expansion.

\chapter{The electrodynamic analogy}

Consider $1+1$ dimensional quantum electrodynamics coupled to a
background
dilaton field $\f$.  The gauge invariant action is
$$
I =  \int d^2 x \>\bigl[ i \overline{\p} \g^\m ( \pa_\m + i A_\m) \p
-\for  e^{-2\f(x)} F_{\m\n} F^{\m\n} \bigr] \ .
\eqn\tenth
$$
The dilaton field is a static non-dynamical background and its only
role in our model is to define a position-dependent coupling
constant,
$$
g^2 (x) = e^{2 \f (x)} \ .
\eqn\eleventh
$$
We will choose a particular dilaton background motivated by the
``linear dilaton vacuum" of $1+1$ dimensional gravity,
$$
\f (x) = - \xw \ ,
\eqn\twelfth
$$
where $\xw$ is the space-like coordinate in Minkowski space.  By
analogy with the black hole case we shall consider the region $\xw
\rightarrow + \infty$ as asymptotic exterior space.  In this region
the coupling $g^2 (\f )$ vanishes exponentially and free fermions can
propagate.  The region $\xw \rightarrow
- \infty$, where the coupling diverges, is analogous to the infinite
throat deep in the interior of certain extreme magnetically charged
black holes [\gidstro].  The question we want to address is whether
or not quantum information is ever lost to an observer at $\xw
\rightarrow + \infty$.  More specifically: is the \smatrix\ for the
asymptotic states at $\xw \rightarrow + \infty$ unitary?

Consider the Penrose diagram in figure~2 for flat $1+1$ dimensional
space-time with a linear dilaton background.  An incoming particle
originating on $\scrimr$ can either propagate to $\scripr$, thereby
escaping the region of strong coupling, or it can continue
propagating toward $\scripl$, in which case it is ``lost'' to the
outside observer.  The unitarity of the \smatrix\ will therefore in
general require asymptotic states to be defined on both $\scripml$
and $\scripmr$.

In both linear dilaton electrodynamics and $1+1$ dimensional dilaton
gravity, left- and right-moving modes of matter fields are uncoupled
at the classical level and in perturbation theory.  In dilaton
gravity this is apparent in the conformal gauge \second\ where the
matter fields $f_i$ satisfy free wave equations.  Incoming
(left-moving) perturbations experience no scattering and the same is
true of right-moving perturbations.  In linear dilaton
electrodynamics the analogous gauge choice is light-cone gauge
$A_-=0$ (or $A_+=0)$, where the Dirac equation,
$$
\g^\m \bigl( \pa_\m + i A_\m \bigr) \, \p = 0 \ ,
\eqn\thirteenth
$$
separates into a pair of uncoupled equations,
$$\eqalign{
\pa_- \p_L =& 0 \ ,   \cr
\bigl( \pa_+ + i A_+ \bigr) \, \p_R =& 0   \ . \cr}
\eqn\fourteenth
$$
The left-moving component appears to be completely decoupled (or the
right-moving component in $A_+=0$ gauge).  In perturbation theory the
asymptotic final states will have particles on both $\scripl$ and
$\scripr$ and it seems that information is inevitably lost to an
observer at $\xw \rightarrow + \infty$.

In both theories, non-perturbative effects associated with quantum
anomalies invalidate the above reasoning.  In dilaton gravity, the
conformal anomaly is responsible for the emission of right-moving
Hawking radiation when a left-moving particle creates a black hole
[\chrful,\cghs].  In linear dilaton electrodynamics the axial anomaly
causes a very similar phenomenon, in which an outgoing current
discharges the field caused by an incoming charged particle, and in
this case one can show that the outgoing radiation carries all the
initial quantum information.

\chapter{Charge hole physics}

\section{Classical solution}

Let us begin with classical $1+1$ dimensional electromagnetism.
Maxwell's equations take the form
$$
\pa_\m \Bigl( {F^{\mu\nu}\over g^2(x)} \Bigr) = j^\n \ .
\eqn\fifteenth
$$
The source-free equations are
$$
\pa_\m \Bigl( {F^{\mu\nu}\over g^2(x)} \Bigr) = 0 \ .
\eqn\sixteenth
$$
In two space-time dimensions the field strength tensor only has one
independent component,\foot{Our conventions are $\e^{01}=+1$ and
metric signature $(-,+)$.}
$$
F^{\m\n} =  F \e^{\m\n} \ ,
\eqn\seventeenth
$$
and we see from \sixteenth\ that ${F\over g^2}$ is constant.  Thus
the general source-free solution is described in terms of one free
parameter $q$,
$$\eqalign{
F^{\m\n} =& q\, g^2(x)\,\e^{\m\n}  \cr
=& q\, e^{-2 \xw}\, \e^{\m\n}      \cr
=& q\, e^{(\xm -\xp)}\, \e^{\m\n} \ ,   \cr}
\eqn\eighteenth
$$
where we have introduced the light-cone coordinates $x^{\pm} = \xz
\pm \xw$.

We will refer to the classical object described by \eighteenth\ as a
``charge-hole''.  It corresponds to a static black hole in dilaton
gravity.  The parameter $q$ which replaces the mass of a black hole
is of course the charge carried by the charge hole.  The analog of
the gravitational collapse solution \sixth\ is a charge hole formed
by an incoming charged particle.  Let the trajectory be $\xp =
\xp_0$, where $\xp_0$ is a constant.  The resulting field is given by
$$
F^{\m\n} = q \, \t (\xp - \xp_0) \,  e^{\xm -\xp} \, \e^{\m\n} \ .
\eqn\twentyfirst
$$
 From Maxwell's equations \fifteenth\ we see that the field in
\twentyfirst\ corresponds to a current
$$
j_+ = q \, \d \bigl( \xp - x_0 ^+ \bigr) \ ,
\eqn\twentysecond
$$
The charge hole vector potential is easily computed in the light-cone
gauge $A_- = 0$. It is given by
$$
A_+(x) = -{q\over 2}
\bigl[\t \bigl(\xp -\xp_0 \bigr) e^{(\xm -\xp)} +\a(\xp)\bigr] \ ,
\eqn\twentythird
$$
where $\a(\xp)$ is arbitrary.

\section{Analogue of Hawking Radiation}

It has been remarked that Hawking radiation can be viewed as pair
production near the event horizon with one particle escaping to
infinity and its partner
falling into the black hole.  This phenomenon also occurs in the
field of a charge hole, where one member of the pair is attracted and
the other is repelled.  The radiation is in the form of charged
particles and, much as in the black hole case, it persists
indefinitely unless back-reaction on the charge-hole is accounted
for.  Apparently, an outside observer only detects the outgoing
particles and must use a density matrix description of the
evaporation process.

The Hawking effect appears in the quantum theory of matter in the
curved, but classical, geometry of a black hole.  Let us therefore
consider the behavior of the quantized fermion field in the
background of a charge-hole.  The gauge field has an effect on the
fermion system through the axial anomaly. The
most efficient way to account for the anomaly is to bosonize the
fermion field.
We therefore begin by reviewing the standard bosonization rules.

One makes the following identifications between fermion variables and
composite operators of a real boson field $Z$:
$$\eqalign{
\overline{\p} \g^\m \p = j^\m &\leftrightarrow \wosp \e^{\m\n}\pa_\n
Z\ ,  \cr \p_L &\leftrightarrow :\exp(i\sqrt{4\pi} Z_L ): \ , \cr
\p_R &\leftrightarrow :\exp(i\sqrt{4\pi} Z_R ): \ , \cr}
\eqn\twentyfive
$$
where we have divided $Z$ into left- and right-moving parts,
$$
Z_{L,R} = {1\over 2} \bigl[Z \mp \int_{\xw}^{\infty} d\xw \, (\pa_0
Z)\bigr]\ .
\eqn\twentysix
$$
Written in terms of the bosonic field the action \tenth\ becomes
$$
I = \int d^2 x \bigl[-\half \pa^\m Z \pa_\m Z - {1\over \sqrt{4\pi}}
\e^{\m\n} F_{\m\n} Z - {1\over 4g^2(x)} F^{\m\n} F_{\m\n}] \ .
\eqn\twentyeight
$$
The equation of motion for $Z$ is
$$
\nabla^2  Z = {1\over \sqrt{4\pi}} \, \e^{\m\n} F_{\m\n} \ ,
\eqn\thirty
$$
which in the background of \twentyfirst\ becomes
$$
\pa_+ \pa_- Z ={q\over 2\sqrt{4\pi}}\, \t (\xp -\xp_0)\, e^{\xm -
\xp} \ .
\eqn\thirtyone
$$
The solution with appropriate boundary conditions corresponding to no
incoming radiation is
$$
Z = - {q\over 2\sqrt{4\pi}} \bigl[ e^{(\xm - \xp)} - e^{(\xm -
\xp_0)} \bigr] \, \t \bigl( \xp - \xp_0 \bigr) \ .
\eqn\thirtytwo
$$
To examine the outgoing radiation we go to the limit $\xp \rightarrow
+ \infty$
$$
Z \rightarrow {q\over 2\sqrt{4\pi}}\,  e^{(\xm - \xp_0)} \ .
\eqn\thirtythree
$$
Using \twentyfive\ we see that an outgoing flux of charge is produced
$$
j_- = {q\over 4\pi} e^{\xm } e^{-\xp_0} \ .
\eqn\thirtyfour
$$
This flux is the analogue of the outgoing Hawking radiation which is
produced by a gravitational collapse.  According to \thirtyfour\ the
radiation persists forever, eventually radiating an infinite charge,
just as the black hole radiates an infinite mass unless back-reaction
is accounted for.

\chapter{\tH -type \smatrix\ for linear dilaton electrodynamics}

In this section we will derive an approximate expression for the
\smatrix\ .  The arguments parallel \tH\ 's construction for black
hole physics as in section~2.

Let us consider the theory in the gauge $A_- = 0$.  The vector
potential describing the field of an infalling charge is given by
\twentythird .  The
right-moving field
$\p_R$ satisfies
$$
\bigl( \pa_+ + i A_+ \bigr) \, \p_R = 0 \ ,
\eqn\thirtyfive
$$
with the general solution
$$
\p_R = \exp [ i S(x) ] \, \x_R \ ,
\eqn\thirtysix
$$
where $S(x) = \int_{-\infty}^{x_+} d\xp \,  A_+ $
and $\x_R$ is a free field.  Thus the effect of the gauge field is to
multiply the outgoing fermion field by a position-dependent phase
factor $e^{i S(x)}$.  Inserting the charge hole vector potential
\twentythird\ gives
$$
S(x) = {q\over 2} \t (\xp - \xp_0) \, \bigl[ e^{(\xm - \xp)} -
e^{(\xm - \xp_0)} \bigr] - {q\over 2} \int \a(\xp) \ .
\eqn\thirtyeight
$$
The second term is an arbitrary constant $c$ times $-q$.  To compute
the \smatrix\ we consider the limit  $\xp \rightarrow + \infty$ ,
where
$$
S(x) \rightarrow - {q\over 2} \bigl[ e^{(\xm - \xp_0)} + c \bigr] \ .
\eqn\thirtynine
$$
Thus the effect of the charge hole gauge field on the fermion system
is a canonical transformation which multiplies $\p_R$ by a phase
$$
\p_R (\xm) \rightarrow \exp[-i {q\over 2} ( e^{\xm - \xp_0} + c ) ]
\, \p_R\ .
\eqn\forty
$$
The transformation \forty\ is a unitary transformation equivalent to
the action
of the unitary operator
$$\eqalign{
U =& \exp \bigl[i \int d\xm \, S(\xm) \,  j_R (\xm)  \bigr]  \cr
  =& \exp \bigl[ {-i \int d\xm \, {q\over 2} \bigl( e^{(\xm - \xp_0)}
+ c \bigr) \, j_R (\xm)}\bigr] \ ,   \cr}
\eqn\fortyone
$$
where
$$
j_R = \p_R^{\dagger} \p_R \ .
\eqn\fortytwo
$$

Let us next suppose that instead of a single delta-function the
incoming charge
is described by a continuous classical flux $j_L(\xm)$.  The
resulting unitary
operator is easily computed to be
$$
U = \exp \bigl[ -{i\over 2} \int d \xp_0 d \xm \, j_L (\xp_0) \,
\bigl( e^{\xm - \xp_0} + c \bigr)
\, j_R ( \xm )
 \bigr]    \ .
\eqn\fortythree
$$
At this point $j_L (\xp) $ is the classical incoming current and
$j_R(\xm)$ is the quantum operator $\p_R^{\dagger} \p_R$.  The
symmetry of the expression, however, suggests that $j_L$ and $j_R$
can be treated on an equal footing as operators in the incoming and
outgoing Fock spaces.

The \smatrix\ \fortythree\ is quite similar to \tH 's gravitational
\smatrix\ \ninth .  In particular, it cannot be a fully correct
description of the scattering any more than \ninth\ is.  To see this,
consider an incoming current $j_L(\xp)$.  According to \fortythree\
the resulting final state is given by
$$
U \ket{0} = \exp \bigr[-{i\over 2} \int  d \xm \, \bigl( A e^{\xm}  +
B c \bigr) \, j_R
(\xm)
\bigr] \ket{0} \ ,
\eqn\fortyfour
$$
where $A$ and $B$ are two moments of $j_L(\xp)$
$$\eqalign{
A =& \int d\xp \, j_L(\xp) e^{-\xp}  \cr
B =&  \int d\xp \, j_L(\xp)    \ .   \cr}
\eqn\fortyfive
$$
Evidently, the final state depends on only two parameters describing
the incident particles.  There is clearly no way that such a final
state can keep track of the full complexity of the incident state and
thus \fortythree\ cannot define a unitary \smatrix\ in the Fock
spaces of in and out particles.

\chapter{Exact \smatrix\ for linear dilaton electrodynamics}

\section{One-particle \smatrix }

Using the bosonization rules of section~4, the action for linear
dilaton electrodynamics can be written
$$
I = \int d^2 x \bigl[ -\half \pa_\m Z \pa^\m Z - {1\over
\sqrt{4\pi}}\, Z\, \e^{\m\n}F_{\m\n} - {1\over 4g^2(x)} F^{\m\n}
F_{\m\n} \bigr]  \ .
\eqn\fortysix
$$
The vector potential can be integrated out to give the following
effective action for the boson field $Z$:
$$
I = \int d^2 x \bigl[ - \half \pa_\m Z \pa^\m Z - {g^2 (x) \over 2
\pi} Z^2
\bigr] \ .
\eqn\fortyseven
$$
This procedure is analogous to that used in [\lslt] to make local the
conformal anomaly term in dilaton gravity.

The $Z$ field now has a mass which increases indefinitely in the
negative $\xw$-direction.  Thus it is evident that any finite-energy
configuration must be totally reflected.  An observer at $\xw
\rightarrow + \infty$ will recover all information.  This fact is not
at all apparent in the original fermionic formulation. Nevertheless,
one can construct a unitary \smatrix\ for fermions.  We will first
illustrate this by computing the amplitude for a single fermion to be
elastically reflected.

An initial state of definite energy is described on $\scrimr$ by
$$
\ket{in} = \int d\xp \, e^{-i p_+ \xp} \p_L(\xp) \, \ket{0} \> ,
\eqn\fortyeight
$$
where $\ket{0}$ is the in-vacuum.  Using the bosonization
prescription
\twentyfive\ this can be written as
$$
\ket{in} = \int  d\xp_0 \, e^{-i p_+ \xp_0} :  e^{ i \sqrt{4\pi}
Z_L(\xp_0) } :  \ket{0} \> .
\eqn\fortynine
$$
 From the boson point of view this is a linear superposition of
coherent states
$$: e^{i \sqrt{4\pi} Z_L(\xp_0) } :  \ket{0} \> .
\eqn\fifty
$$
Each such coherent state is identified with a classical configuration
$Z_C(x)$
and evolves in time into another coherent state according to the
classical
equations of motion.  The initial configuration corresponding to
\fifty\ is a
left-moving step function
$$
Z_C = \sp \t ( \xp - \xp_0 ) \ .
\eqn\fiftyone
$$
Note that the charge carried by a configuration is given by
$$
Q = \int_{-\infty}^{+\infty} dx \, j^0
= \int_{-\infty}^{+\infty} dx \, \wosp {\pa Z\over \pa x}
= \wosp \bigl[Z_C(+\infty) - Z_C(-\infty)\bigr] \ .
\eqn\fiftytwo
$$
Thus the net incoming charge is proportional to the height of the
step function.

The incoming state has the form \fiftyone\ on  $\scrimr$ i.e. at $\xm
\rightarrow -\infty$.  To find the subsequent evolution we need to
solve the classical equations for $Z_C$,
$$
\pa_+ \pa_- Z_C = - {1\over 4\pi} g^2(x) Z_C \> = - {1\over 4\pi}
e^{(\xm - \xp)} Z_C \ ,
\eqn\fiftythree
$$
subject to the boundary conditions \fiftyone\ at  $\scrimr$.  Note
that we do not need to impose boundary conditions on $\scriml$
because the mass term in \fiftythree\ diverges there forcing $Z$ to
vanish.  The appropriate solution can for example be found by using a
a coordinate system which turns \fiftythree\ into a Klein-Gordon
equation with a uniform tachyonic mass~[\alstrom].  It is given by
$$
Z_C =\sp \, \t (\xp - \xp_0)\, J_0 \bigl[ \wosp e^{\half \xm} \sqrt{
e^{-\xp_0} - e^{-\xp}}  \bigr]  \ .
\eqn\fiftyfour
$$

It is instructive to examine \fiftyfour\ on a series of time slices,
showing how the field evolves.  This is illustrated in figure 3.  We
see that the point charge continues to penetrate toward $\xw
\rightarrow - \infty$ but becomes more and more tightly screened as
time evolves.  Asymptotically it becomes totally screened.  A
reflected charge of equal magnitude moves off to the right towards
the asymptotic weak coupling region and it is followed by a series of
pairs with ever higher frequency but lower charge.  The degenerate
left-moving ``blip" is an artefact of having arbitrarily high-energy
components in the localized state $\p(x_0) \ket{0}$.  The actual
initial state \fortyeight\ is a superposition of such
localized states and has finite energy.

The asymptotic out-state on $\scripr$ is obtained by taking the limit
$\xp \rightarrow + \infty$ in \fiftyfour\
$$
Z_C \rightarrow \sp J_0 \bigl[ \wosp e^{\half (\xm -\xp_0)}\bigr]\ .
\eqn\fiftyfive
$$
The corresponding coherent quantum state is given by
$$
: \exp \bigl[ i\int d\xm \, 2\pa_- Z_C(\xm, \xp_0) \, Z_R(\xm)
\bigr] :  \ket{0} \> .
\eqn\fiftysix
$$
Thus the final state is
$$
\int d\xp_0 \, e^{-i p_+ \xp_0} : \exp \bigl[ i\int d\xm \, 2\pa_-
Z_C(\xm, \xp_0) \, Z_R(\xm)   \bigr] :  \ket{0} \> .
\eqn\fiftyseven
$$
The elastic scattering amplitude is the overlap of this state with an
outgoing fermion,
$$
\int dx_0^-  e^{ i q_- \xm_0}  \bra{0}  : \exp[-i \sqrt{4\pi}
Z_R(x_0^-)] : \ .
\eqn\fiftyeight
$$
A standard coherent state calculation yields an amplitude,
$$\eqalign{
A(q_-, p_+) =& \int dx_0^+ dx_0^-  e^{i(q_- \xm_0 - p_+ \xp_0)} \,
\exp \bigl(\int {dv \over v + i\epsilon} J_0 \bigl[ \wosp e^{\half (v
+ \xm_0 -\xp_0)} \bigr] \,  \bigr) \cr
=& 2\pi\, \delta (p_+{-}q_-) \int dx e^{-i p_+ x} \exp \bigl(\int {dv
\over v + i\epsilon} J_0 \bigl[ \wosp e^{\half (v - x)} \bigr] \,
\bigr) \ .  \cr}
\eqn\fiftynine
$$
The $i\epsilon$ prescription takes care of the ultra-violet
divergences but, as it stands, this expression is still infra-red
divergent.  This is because we have used a simple logarithm for the
boson propagator in the coherent state calculation, whereas a more
careful evaluation, using a regularized propagator, would give a
finite result.  An alternative, if somewhat crude, subtraction
procedure is simply to subtract from the Bessel function in the $v$
integral in \fiftynine\ a step function $\theta (v_0 - v)$, which
cancels the $v\rightarrow -\infty$ infrared divergence.  The
dependence on the subtraction point, $v_0$, can be absorbed into the
overall normalization of the amplitude, which we have not kept track
of here.  If desired, the normalization can be determined by the
physical requirement that the probability for elastic reflection of a
fermion approaches unity as the energy tends to zero.

\section{The full \smatrix}

Now we want to construct the full operator \smatrix\ for the
scattering of arbitrary fermion states.  The best way to achieve this
is to first obtain the exact \smatrix\ for bosons and then appeal to
the equivalence between the Hilbert spaces of the bosons and fermions
to read off the fermion \smatrix .
The boson amplitudes are easy to obtain because \fortyseven\ defines
a free field theory.  Let us start with the LSZ-reduced expression
for a one-particle \smatrix\ element for bosons, which is obtained by
sandwiching the operator
$$
S_{1\rightarrow 1} = i \int d^2x_1 \,d^2x_2 \, Z_R(x^-_1) \,
\overrightarrow \nabla^2_1 G(x_1,x_2) \overleftarrow \nabla^2_2 \,
Z_L(x^+_2)
\eqn\sixty
$$
between asymptotic single boson Fock states.  By using the coordinate
system in which the equation of motion for $Z$ becomes a tachyonic
Klein-Gordon equation, and demanding that the propagator vanishes in
the strong coupling region, one is led to
$$
G(x_1,x_2) = \sp J_0\bigl[\wosp \sqrt{\vert e^{-x^+_1}-
e^{-x^+_2}\vert \> \vert e^{x^-_1}-e^{x^-_2}\vert}\bigr]\ .
\eqn\sixtyone
$$
After inserting this propagator into \sixty\ and some integrations,
we find
$$
S_{1\rightarrow 1} = i \int dx^-_1 dx^+_2 \, \partial_- Z_R(x^-_1) \,
J_0 \bigl[\wosp e^{{1\over 2}(x^-_1 - x^+_2)} \bigr] \, \partial_+
Z_L(x^+_2) \ .
\eqn\sixtytwo
$$
For a free field theory the full \smatrix\ is obtained by
exponentiating the single particle expression
$$
S = \exp \Bigl[i \int dx^-_1 dx^+_2 \, \partial_- Z_R(x^-_1)
\, J_0 \bigl[\wosp e^{{1\over 2}(x^-_1 - x^+_2)} \bigr] \,
\partial_+ Z_L(x^+_2)\Bigr] \ .
\eqn\sixtythree
$$

Exactly the same operator expression can now be used to compute
\smatrix\
elements in the fermion basis.  For example, the single-particle
matrix element
\fiftynine\ is given by
$$
\int d\xp d\xm \, e^{i(q_- \xm - p_+ \xp)}\, \bra{0} : e^{-i
\sqrt{4\pi} Z_R(\xm)} : S :e^{+i \sqrt{4\pi} Z_L(\xp)} : \ket{0} \ .
\eqn\sixtyfour
$$
The general expression \sixtythree\ can be written directly in
fermion language by using the fermion-boson correspondence,
$$
\wosp \e^{\m\n} \pa_\n Z = j^\n \ ,
\eqn\sixtyfive
$$
giving
$$
S = \exp \bigl(i\pi \int d\xp \, d \xm \,  j_R(\xm) \,
J_0 \bigl[ \wosp e^{\half
(\xm -\xp)} \, \bigr] j_L(\xp) \, \bigr) \ .
\eqn\sixtysix
$$

Evidently the exact \smatrix\ is of the form advocated  by \tH\ but
with a more
complicated kernel than \fortyfour .  In fact, the correspondence can
be seen directly by expanding the Bessel function in a power series
in $e^{\xm - \xp}$.  The first two terms of the expansion pick up the
moments in  \fortyfour \foot{In fact there is a factor of two
discrepancy between the coefficients in \fortyfour\ and \sixtysix .
This factor can be traced to the asymmetric treatment of incoming and
outgoing currents in section~5 and does not appear in a more
symmetric calculation}.  The full series expansion involves all the
moments making it possible for unitarity to be restored.

The meaning of the higher terms in the series expansion can be given
a graphical interpretation.  Each successive power of $e^{\xm - \xp}$
corresponds to a closed loop of fermions in the gauge field
propagator, which enters into the calculation of the phase shift of
the outgoing fermions.

\chapter{Information retrieval}

Having established the existence of a unitary \smatrix\ for linear
dilaton electrodynamics, it is interesting to ask how the information
in a complex initial state is radiated back.

For example, suppose an initial state of given total charge $Q$
described by a coherent state with some modulations on the $Z$-field.
Now consider boosting the configuration to higher energy so that the
information carrying modulations are squeezed into a smaller volume.
At extremely high energy it will become indistinguishable from a step
function whatever its initial profile.  However, boosting a
configuration cannot change its information content.  How, then, does
the final state remember the incident structure?

The answer is in the very high-frequency exponentially attenuating
tail in figure~3.  In the limit of infinite boost, the tail extends
to $\xw \rightarrow - \infty$, and because of the increasing
frequency in this region it carries infinite energy.  In a finite
energy configuration, the tail is bounded.  The details of the
initial configuration are coded in the details of the high-frequency
low-amplitude tail.  In other words, an energetic collection of  low
charge fermion pairs trails the main bulk of the outgoing charged
radiation and information about all the details of the boosted
initial state are coded into modulations on that tail.

We do not know to what extent the mechanism for information retrieval
carries over to two-dimensional gravity, let alone the real world.
Obviously we cannot expect the information in a black hole to be
radiated in a late tail of high energy quanta since most of the
energy of the black hole will already have been radiated.  Note,
however, that in the analogy between two-dimensional gravity and
linear dilaton electrodynamics, gravitational energy is replaced by
electric charge.  The information carrying tail in linear dilaton
electrodynamic carries very little charge which should perhaps be
interpreted in gravity as information escaping from a black hole
remnant in a long tail of very soft radiation, containing a large
number of quanta.  Since the coding of the information into long
wavelength quanta would have to be a very slow process [\hawking,\ac]
such a proposal would probably suffer from the drawbacks of stable
remnant theories.  We hope to return to these points.

Another point worth noting is that the unitarity of the \smatrix\
depends on the field content of the theory.  For example, if two
species of fermions were coupled to the electromagnetic field the
difference of their charge densities would not be expelled from the
strongly coupled region.  In this case one linear combination of the
bosonizing fields would carry information to $x^1\rightarrow -\infty$
where it would be lost to an outside observer.  Perhaps information
can only be conserved in some theories.

\vskip 2cm
\noindent
{\undertext{Acknowledgements:}}  The authors would like to thank
S.~Giddings,  J.~Russo and A.~Strominger for useful discussions.

\FIG\figone{The effect of an infalling shock wave on a black hole
geometry.  The event horizon shifts outward.}
\FIG\figtwo{``Penrose diagram" for a charge-hole.}
\FIG\figthree{Evolution in time of the bosonizing field, for an
incoming fermion.}

\refout
\figout
\end